\documentclass{Interspeech}

\interspeechcameraready

\title{Counterfactual Activation Editing for\\ Post-hoc Prosody and Mispronunciation Correction in TTS Models}

\author[affiliation={1}]{Kyowoon}{Lee}
\author[affiliation={1}]{Artyom}{Stitsyuk}
\author[affiliation={2}]{Gunu}{Jho}
\author[affiliation={2}]{Inchul}{Hwang}
\author[affiliation={1,3}]{Jaesik}{Choi}

\affiliation{}{KAIST}{South Korea}
\affiliation{}{Samsung Electronics}{South Korea}
\affiliation{}{INEEJI}{South Korea}
\email{\{leekwoon, stitsyuk, jaesik.choi\}@kaist.ac.kr, \{gunu.jho, inc.hwang\}@samsung.com}

\usepackage{comment}

\usepackage{subcaption}

\begin{document}

\maketitle

\begin{abstract} 
Recent advances in Text-to-Speech (TTS) have significantly improved speech naturalness, increasing the demand for precise prosody control and mispronunciation correction. Existing approaches for prosody manipulation often depend on specialized modules or additional training, limiting their capacity for post-hoc adjustments. Similarly, traditional mispronunciation correction relies on grapheme-to-phoneme dictionaries, making it less practical in low-resource settings. We introduce Counterfactual Activation Editing, a model-agnostic method that manipulates internal representations in a pre-trained TTS model to achieve post-hoc control of prosody and pronunciation. Experimental results show that our method effectively adjusts prosodic features and corrects mispronunciations while preserving synthesis quality. This opens the door to inference-time refinement of TTS outputs without retraining, bridging the gap between pre-trained TTS models and editable speech synthesis.  
\end{abstract}




\section{Introduction}

\let\oldthefootnote\thefootnote

\let\thefootnote\relax
\makeatletter\def\Hy@Warning#1{}\makeatother
\footnotetext{
Audio samples can be found at \href{https://leekwoon.github.io/cae-tts}\url{https://leekwoon.github.io/cae-tts}
}

\let\thefootnote\oldthefootnote
\setcounter{footnote}{0}

Advancements in Text-to-Speech (TTS) models \cite{wang2017tacotron, shen2018natural, li2019neural, ren2019fastspeech, ren2020fastspeech} and neural vocoders \cite{oord2016wavenet, prenger2019waveglow, kumar2019melgan, kong2020hifi} have made synthetic voices nearly indistinguishable from human speech.
Consequently, more attention has been attracted by refining the expressiveness of synthetic voices, particularly through the accurate manipulation of prosodic features and pronunciation.

Traditionally, controlling prosody and correcting mispronunciations in TTS models have involved distinct approaches. For prosody control, one popular strategy has been learning a latent space, from which embeddings are sampled during inference to supplement the text with necessary prosodic information \cite{wang2018style, skerry2018towards, hsu2018hierarchical, vsimko2023prosody}. However, due to its unsupervised nature, selecting the right embedding can be challenging. An alternative strategy involves adding specific layers to the TTS model that are trained to extract and manipulate this prosodic information directly from the input sequence \cite{ren2019fastspeech, ren2020fastspeech, mohan2021ctrl, raitio2022hierarchical}, but this requires extensive preprocessing. Both of these methods, however, lack straightforward applicability to already trained models, such as Tacotron 2 \cite{shen2018natural} or Transformer TTS \cite{li2019neural}, without undergoing a retraining process.

Similarly, the field of pronunciation correction has heavily relied on pronouncing dictionaries to convert grapheme to phoneme. While effective, this method struggles with scalability, precluding TTS models from being useful in low-resource languages. Diversifying pronunciations during training helps, but the skewed word distribution in natural language (Zipfian distribution) \cite{taylor2019analysis} limits its utility. In general, TTS training datasets cannot match the comprehensive coverage of pronunciation dictionaries, highlighting the need for a more flexible and encompassing strategy.

In this paper, we introduce an \emph{orthogonal} framework leveraging \emph{Counterfactual Activation Editing} (CAE) to directly manipulate the prosody and pronunciation in pretrained TTS models. Rather than adding specialized layers or relying on extensive external resources, our method exploits \emph{post-hoc decomposability} by examining and editing intermediate representations after the model is trained. By posing counterfactual questions (e.g.,~``What would the internal representation look like if the model aimed for a higher pitch rather than the current, lower one?''), we perform fine-grained modifications to hidden activations, thereby refining prosodic features and correcting mispronunciations at inference time.

The main contributions of this paper are summarized as follows: We introduce an approach that provides post-hoc adjustability of speech output, allowing for the adaptation of TTS models to new requirements without retraining. Additionally, we demonstrate the effectiveness of the proposed method through rigorous experimental validation.

\section{Related Work}

\subsection{TTS Prosody Control}

Text can be spoken in various ways due to semantic nuances, speaking styles, or inherent variability. Traditional approaches, such as unit-selection, capture this variability through speech databases \cite{strom2006expressive}.
In contrast, recent studies model prosodic variation by predicting key prosodic features such as pitch, duration, and energy from their embedding spaces \cite{ren2019fastspeech, ren2020fastspeech, mohan2021ctrl, bandekar2023speaking}, or by using implicit representations learned from a reference encoder to capture the nuanced variations not specified by text alone \cite{wang2018style, skerry2018towards, hsu2018hierarchical}. 

Controlling intermediate representations has also been considered in prior work through the use of embedding bias, calculated by assessing the extent of translation required to achieve specific modifications in acoustic features within multidimensional scaling coordinates \cite{lenglet2022speaking}. However, such modified representations risk deviating from the data manifold, and their application has been primarily confined to controlling duration.

\subsection{TTS Pronunciation Control}

A critical challenge for end-to-end TTS models, which operate without the aid of grapheme-to-phoneme dictionaries or predictive model, is polyphone disambiguation \cite{zhang2020unified}. For TTS models to accurately convert graphemes into phonemes, their linguistic encoder must internalize the varied pronunciation rules, but fully internalizing them is difficult, leading to inevitable pronunciation errors in synthesized speech.

To mitigate mispronunciations, unit selection concatenates recorded speech fragments from a database. However, this often leads to noticeable join artifacts and requires a single-speaker database, limiting the use of non-target speaker data. 

Recently, a model-centric approach called as the Speech Audio Corrector (SAC) \cite{fong2022speech} has been introduced. It leverages speech codes aligned with words, derived from self-supervised learning models to correct mispronunciations at the word level. In contrast, we introduce a model-agnostic approach that manipulates intermediate representations in TTS models.

\section{Proposed Method}

\begin{figure}[t]
    \centering
    \begin{subfigure}[b]{0.49\columnwidth}
        \centering
        \includegraphics[width=1\linewidth]{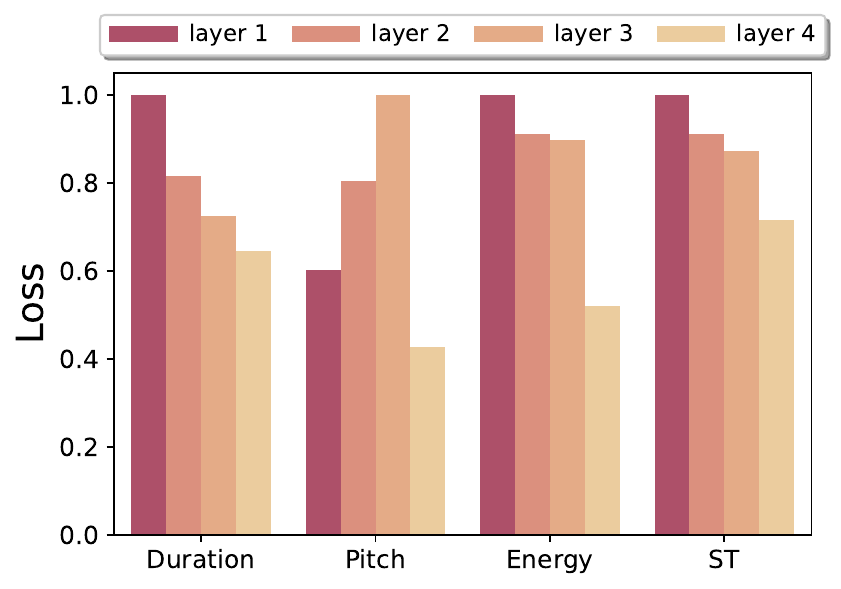}
        \vspace{-0.5cm}
        \caption{Layer-level Analysis}
        \label{fig:layer}
    \end{subfigure}
    \hfill
    \begin{subfigure}[b]{0.49\columnwidth}
        \centering
        \includegraphics[width=1\linewidth]{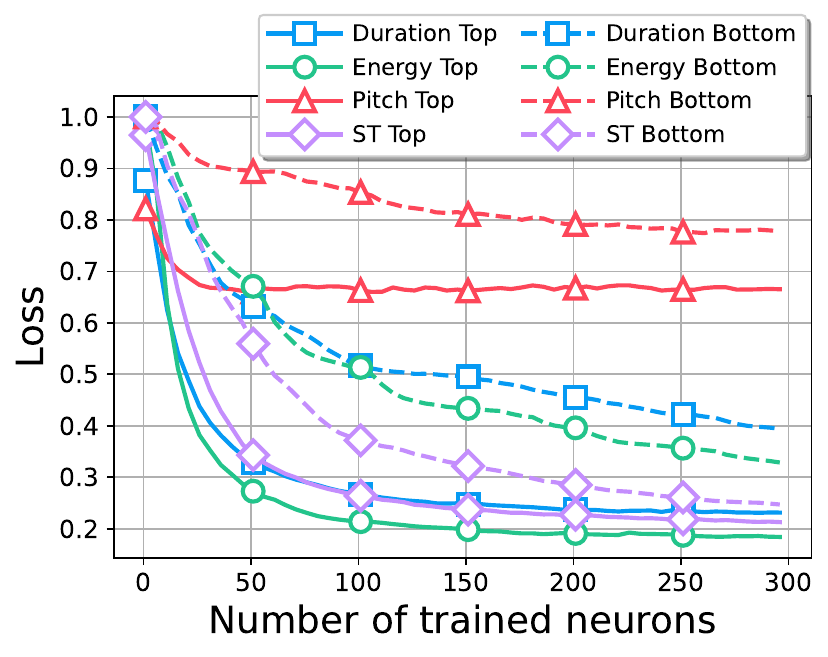}
        \vspace{-0.5cm}
        \caption{Neuron-level Analysis}
        \label{fig:ablated_neurons}
    \end{subfigure}
    \vspace{-0.10cm}
    \caption{Acoustic Correlation Analysis. (a) Classifier prediction losses across Tacotron 2 encoder layers, normalized with the highest value as one. (b) Effect of training only the most or least important neurons on classifier performance.}
    \vspace{-0.50cm}
    \label{fig:acoustic_correlation_analysis}
\end{figure}

We first analyze intermediate representations produced by the encoder of an end-to-end TTS model, applicable to any encoder-decoder architectures.

Consider a sequence of input features $\mathbf{t}=\{\mathbf{t}_1,\dots,\mathbf{t}_n\}$, and let $\mathbb{M}$ be a neural network model that transforms $\mathbf{t}$ into a series of internal representations $\mathbf{t} \xrightarrow{\mathbb{M}} \mathbf{x}=\{\mathbf{x}_1,\dots, \mathbf{x}_n\}$. In this context, $\mathbb{M}$ represents the encoder, with $\mathbf{t}$ as either grapheme or phoneme embeddings. Let $\mathbf{x}_i$ denote the activation in the last layer of encoder for the $i$-th input feature.\footnote{
While our analysis primarily concentrates on neurons in the top layer of the encoder, the approach can be applied to other layers.}

\subsection{Acoustic Correlation Analysis}\label{subsec:acoustic_correlation_analysis}

In the context of a classification task\footnote{Prosody typically suits regression due to its continuous nature; however, for clarity, we discuss it under classification contexts here.}, we aim to predict a specific property, denoted as $\mathbf{l}_i$, from a set of properties $\mathcal{P}$ presumed to be inherently learned by the model $\mathbb{M}$, for example, prosodic or pronunciation features in TTS model. Our objective is to examine intermediate representations within $\mathbb{M}$ to assess their significance in recognizing the property $\mathbf{l}_i \in \mathcal{P}$.

For analyzing prosody properties such as duration, pitch, and energy, we first extract durations using an automatic speech recognition (ASR) system with connectionist temporal classification (CTC) alongside a backtranslation loss inspired by \cite{lux2023exact}. This system facilitates the extraction of precise alignments by sorting the posterior probabilities and applying monotonic alignment search (MAS). Subsequently, we calculate the average pitch and energy values based on these alignments to provide a comprehensive analysis.

For pronunciation analysis, we capture discrete pronunciation patterns at the semantic token (ST) level using the layer-6 representation of HuBERT \cite{hsu2021hubert}, which operates at a 50Hz frame rate for 16kHz audio, to generate a sequence of semantic representations. These representations are aligned to input features using the same technique as for prosody, then aggregated and discretized into semantic tokens via k-means clustering (100 clusters). This process helps identify unique pronunciation patterns, supported by previous studies \cite{fong2022speech, hsu2021hubert, ji2022predicting} that show HuBERT encodes phonetic information.

We train a logistic regression model on pairs of intermediate representations and their labels ${\mathbf{x}_i, \mathbf{l}_i}$, using cross-entropy loss. Previous studies have demonstrated that in the analysis of neural network representations, the performance characteristics of non-linear models closely resemble those of linear models \cite{belinkov2017neural, dalvi2019one}. Our analysis reveals that TTS model representations are densely packed with information critical to both prosody and pronunciation, as shown in Figure \ref{fig:acoustic_correlation_analysis}.

\subsection{Counterfactual Activation Editing}\label{subsec:cae}

To manipulate the internal representations in a way that changes a predicted property, we consider a counterfactual question: for instance, ``What would the internal representation look like if the synthesized speech had a higher pitch rather than a lower pitch?''. To address this, we introduce \emph{Counterfactual Activation Editing (CAE)}, which manipulates internal representations to achieve the desired property.



Let $f: \mathcal{X} \to \mathbb{R}^C$ be a classifier that assigns to an activation vector $\mathbf{x}_i \in \mathcal{X}$ the probability of belonging to a class \( c \in \{1, \ldots, C\} \). 
For a given step size \(\eta\) and a target class \(c\), CAE updates the activation via gradient ascent:
\begin{align}
\label{eq:cf_x}
\mathbf{x}_i^{(k+1)}=\mathbf{x}_i^{(k)} + \eta\frac{\partial f_c}{\partial\mathbf{x}_i}(\mathbf{x}_i^{(k)}).
\end{align}
This process continues until the classifier output for the target class \(c\) exceeds a chosen threshold.\footnote{In a regression setting, where no explicit class boundaries exist, one may directly increase or decrease the regressor output toward the desired value.} The procedure can be formulated as:
\begin{align}
\max_{\mathbf{x'}_i \in N(\mathbf{x}_i^{(k)}) \subset \mathbb{R}^d} f_c(\mathbf{x'}_i),
\end{align}
where $N(\mathbf{x}_i^{(k)})=\{\mathbf{x} \in \mathbb{R}^d \mid d(\mathbf{x}, \mathbf{x}_i^{(k)}) < r\}$ denotes a neighborhood of radius \(r\) around the current activation \(\mathbf{x}_i^{(k)}\), which depends on the optimization step size~\(\eta\).

\subsubsection{Manifold Preserving CAE}
In practice, the activation $\mathbf{x}_i$ lies in a much lower-dimensional space, such as a $k$ dimensional manifold $\mathcal{M}$, where $k \ll d$. The naive gradient-ascent procedure in the ambient space \(\mathbb{R}^d\) may move \(\mathbf{x}_i\) off the manifold, introducing unstructured noise rather than meaningful, semantically coherent changes. To mitigate this, we constrain updates to a local tangent space:
\begin{align}
N_{\mathcal{T}}(\mathbf{x}_i^{(k)})=\{\mathbf{x} \in \mathcal{T}_{\mathbf{x}_i^{(k)}}\mathcal{M} \mid d(\mathbf{x}, \mathbf{x}_i^{(k)}) < r\},
\end{align}
where \(\mathcal{T}_{\mathbf{x}_i^{(k)}}\mathcal{M}\) is the tangent space of \(\mathcal{M}\) at \(\mathbf{x}_i^{(k)}\). We approximate these tangent spaces using an autoencoder, leveraging the information bottleneck to learn the manifold geometry~\cite{joshi2019towards, dombrowski2023diffeomorphic}.

Concretely, we first map \(\mathbf{x}_i\) to the latent space \(\mathbf{z}_i\) of a \(\beta\)-VAE~\cite{higgins2016beta}. We then perform gradient ascent in \(\mathbf{z}_i\), ensuring that the updates remain close to the data manifold. Formally,
\begin{align}
\mathbf{z}_i^{(k+1)} = \mathbf{z}_i^{(k)} + \eta \frac{\partial (f \circ g)_c}{\partial \mathbf{z}_i}(\mathbf{z}_i^{(k)}),
\end{align}
where \(g\) is the decoder mapping from the latent space \(\mathcal{Z}\) back to the original activation space \(\mathcal{X}\). By constraining edits within \(\mathcal{Z}\), the resulting activations lie closer to the manifold \(\mathcal{M}\), yielding counterfactual changes that are semantically meaningful.

\subsubsection{Prototype Loss}



Even with manifold preserving updates, the latent space of a $\beta$-VAE may include multiple subregions corresponding to different pronunciations or prosodic patterns. Consequently, large or unconstrained edits could inadvertently shift the activation into a subregion that alters the intended content. To address this, we incorporate a \emph{prototype loss} that anchors the edited latent vectors to a representative code within a Vector Quantized VAE (VQ-VAE)~\cite{van2017neural}.

Let $\text{Enc}(\cdot)$ and $\text{Dec}(\cdot)$ be the encoder and decoder of the VQ-VAE, and let $\mathbf{e} \in \mathbb{R}^{K \times D}$ denote the learned codebook with $K$ discrete embeddings of dimension $D$. Given a latent representation $\mathbf{z}_i$ obtained via our Manifold Preserving CAE process, we find its nearest codebook vector $\mathbf{e}_k$:
\begin{align}
\mathbf{e}_k 
&= 
\text{Quantize}(\text{Enc}(\mathbf{z}_i)) \nonumber \\ 
&=
\underset{\mathbf{e}_j}{\arg\min}||\text{Enc}(\mathbf{z}_i) - \mathbf{e}_j||_2.
\end{align}
Decoding $\mathbf{e}_k$ yields the \emph{prototype} representation:
\begin{align}
\text{proto}(\mathbf{z}_i)=\text{Dec}(\mathbf{e}_k)
\end{align}
To ensure edits remain faithful to the intended pronunciation, we augment the existing objective with a prototype loss that encourages $\mathbf{z}_i$ to remain close to its prototype:
\begin{align}
\mathcal{L}_{proto} = \alpha \cdot ||\mathbf{z}_i - \text{proto}(\mathbf{z}_i)||_2^2.
\end{align}
where $\alpha$ is a weighting factor. By enforcing proximity to a prototype in the codebook, this loss discourages transitions to latent subregions that would alter the target speech content.

\section{Experimental Setup}

\begin{figure}
    \centering
    \includegraphics[width=0.99\linewidth]{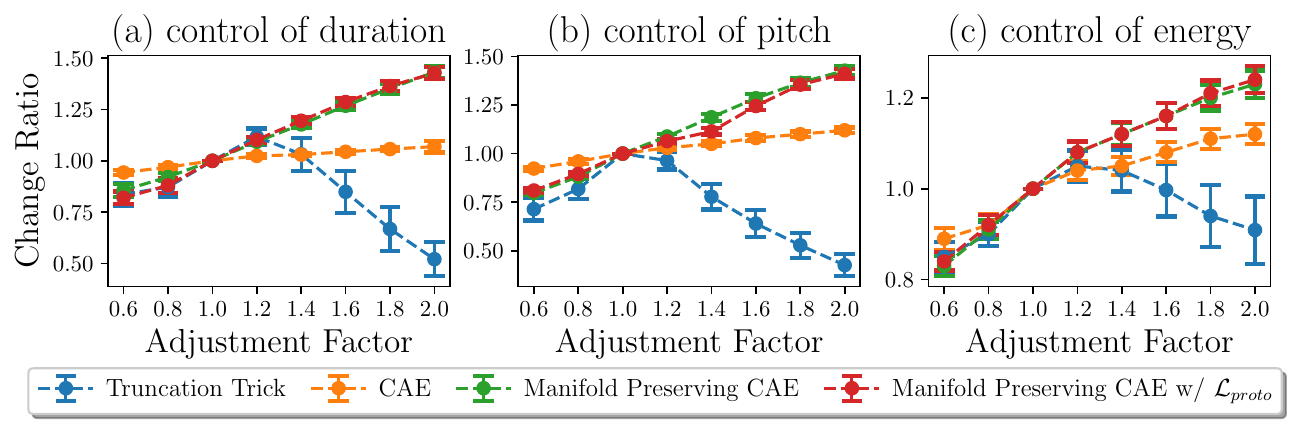}
    \vspace{-0.2cm}
    \caption{Ratio of change in prosodic features. Values above 1 indicate an increase, while values below 1 indicate a decrease.}
    \vspace{-0.3cm}
    \label{fig:prosody_control}
\end{figure}

\subsection{TTS Model}

We experimented with Tacotron 2 \cite{shen2018natural} encoder-decoder architecture, a model with about 28 million parameters, including an encoder with three convolutional layers followed by an LSTM layer. Although Tacotron~2 generates highly natural speech, it lacks explicit mechanisms to capture diverse prosodic contours without additional modules or training data. This limitation makes Tacotron~2 an ideal testbed for our \emph{post-hoc} modification approach, as any improvement in controllability can be primarily attributed to our method rather than inherent model capacity. Phoneme inputs are employed to control prosody, whereas grapheme inputs are used for pronunciation adjustments, to better represent resource-limited scenarios. We base our Tacotron 2 on NVIDIA's implementation\footnote{\href{https://github.com/NVIDIA/tacotron2}{https://github.com/NVIDIA/tacotron2}} and use a pre-trained WaveGlow \cite{prenger2019waveglow} as the vocoder.

\begin{figure}
    \centering
    \includegraphics[width=0.99\linewidth]{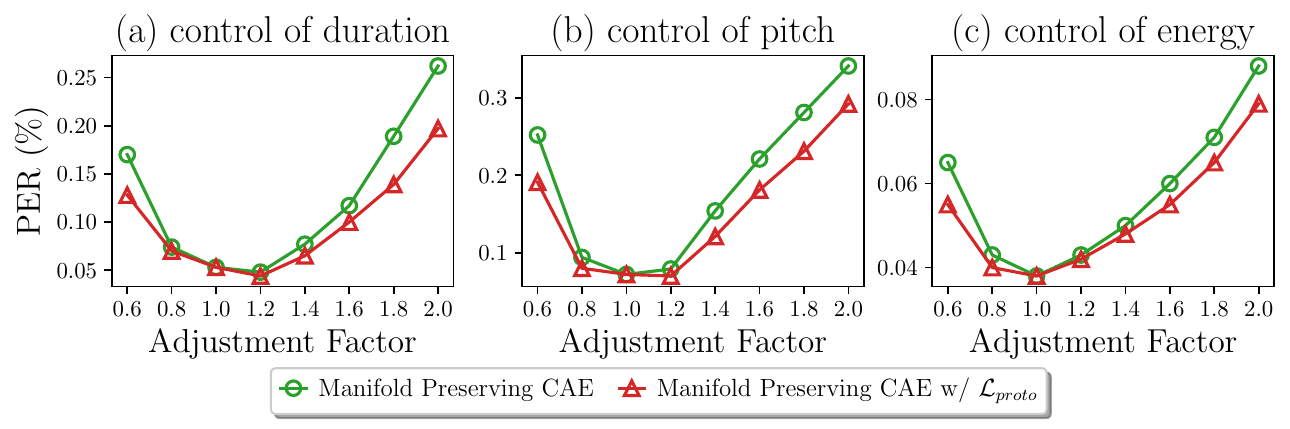}
    \vspace{-0.2cm}
    \caption{Phone-Error-Rate (PER) comparison with and without prototype loss, measured using the Whisper ASR system.}
    \vspace{-0.4cm}
    \label{fig:prosody_control_per}
\end{figure}

\subsection{Dataset}

For training Tacotron 2, we use the LJSpeech dataset \cite{ljspeech17}, containing 13,100 English audio clips (around 24 hours) from a single female speaker, along with text transcripts. The dataset was split into 12,500 samples for training, 500 for validation, and 100 for testing.

\subsection{Training}
In the training phase of the classifier, we extracted acoustic properties and semantic tokens as detailed in Section \ref{subsec:acoustic_correlation_analysis} from the LJSpeech training dataset.
For prosody prediction, a linear regression model was employed, using mean squared error loss augmented with elastic net regularization, setting the l1 and l2 regularization hyperparameters, $\lambda_1$ and $\lambda_2$, both to 0.001. For semantic token prediction, we used a logistic regression model trained using cross-entropy loss. Both models were trained using a learning rate of $0.01$. Additionally, for the training of the $\beta$-VAE, we set $\beta$ to 0.04 and the dimension of the latent space to 16, with a learning rate of 0.0002.

\section{Results}
Our experiments study the following questions: (1) Are acoustic features encoded by the Tacotron 2 encoder? (2) Can \emph{counterfactual activation editing} (CAE) effectively manipulate prosody? (3) Can CAE further correct mispronunciations?


\subsection{Acoustic Correlation Analysis}

To assess the capacity of Tacotron 2 encoder layers to encapsulate meaningful acoustic features, classifiers were trained using activations from different layers of a pre-trained Tacotron 2 model. Prediction losses across the encoder layers, shown in Figure \ref{fig:layer}, demonstrate that acoustic information is captured at varying levels. Notably, the last LSTM layer is more accurate in capturing acoustic properties compared to the earlier convolutional layers.

In addition, we conduct a detailed examination of individual neurons to identify those that significantly influence certain properties. This involved training only the most or least significant neurons from the last layer of the encoder, ranked by their importance based on classifier weights \cite{dalvi2019one}, and observing the effect on classifier efficacy. Figure \ref{fig:ablated_neurons} shows that classifier performance improves more significantly when top-ranked neurons are trained (solid lines) compared to when the least important neurons are trained (dashed lines), indicating the crucial role of specific neurons in encoding acoustic properties.

\begin{figure}[t]
    \centering
    \begin{subfigure}[b]{0.95\linewidth} 
        \centering
        \includegraphics[width=1\linewidth]{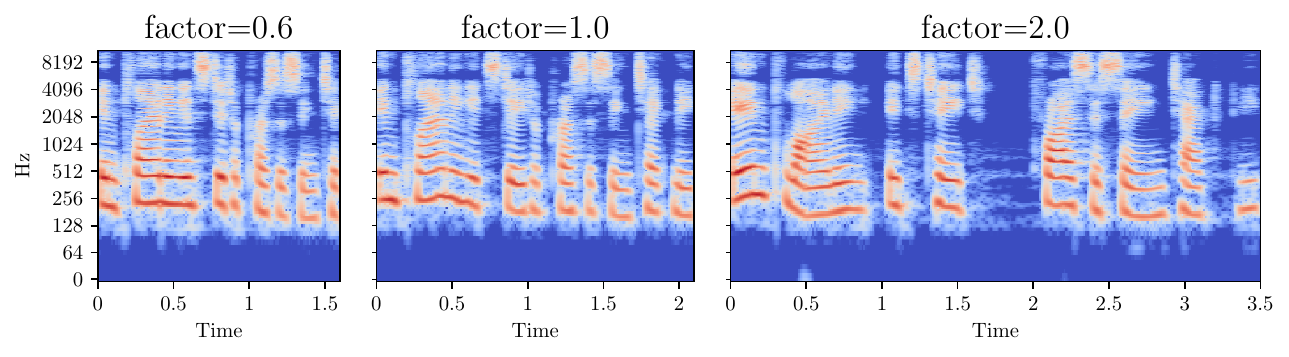}
        \vspace{-0.55cm}
        \caption{Control of Duration}
        \label{fig:dur mel}
    \end{subfigure}
    \\
    \begin{subfigure}[b]{0.95\linewidth}
        \centering
        \includegraphics[width=1\linewidth]{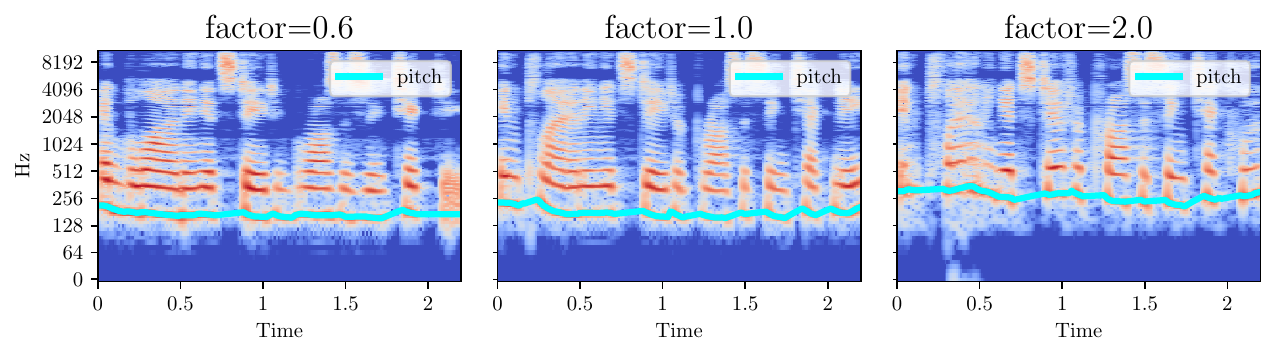}
        \vspace{-0.55cm}
        \caption{Control of Pitch}
        \label{fig:f0 mel}
    \end{subfigure}
    \vspace{-0.1cm}
    \caption{Spectrograms of the synthesized speech using Manifold Preserving CAE for prosody control. The input text used for synthesis is ``In planning its data processing techniques.''}
    \vspace{-0.4cm}
    \label{fig:prosody_mel}
\end{figure}
\subsection{Prosody Control}

We investigate \emph{Counterfactual Activation Editing} (CAE) for controlling prosodic features in a pre-trained Tacotron~2 model. We compare three variations: CAE, which edits the encoder activations directly; Manifold Preserving CAE, which first encodes activations into a VAE latent space and edits them there to prevent manifold deviation; and Manifold Preserving CAE with prototype loss, which further constrains edits around representative codebook vectors to preserve intelligibility. We also compare against a truncation trick~\cite{brock2018large}, which shifts activations toward values observed in the training set.

We specifically modify the final-layer encoder activations for each phoneme in order to scale its prosodic property (e.g.,~pitch, duration, or energy) by a predefined adjustment factor. Let $f(\mathbf{x}_i)$ be a regressor predicting a prosodic feature from the activation vector $\mathbf{x}_i$. Suppose $f(\mathbf{x}_i)=p_i$ is the current predicted pitch. To increase or decrease pitch by a factor $\lambda>0$, we adjust $\mathbf{x}_i$ so that $f(\mathbf{x}_i')$ approaches $\lambda\,p_i$, following the gradient-based procedure described in Section~\ref{subsec:cae}.

Figure~\ref{fig:prosody_control} shows the ratio between the modified prosodic feature and its original value, where values above 1 indicate an increase and values below 1 indicate a decrease. Manifold Preserving CAE consistently achieves more precise prosodic changes compared to both CAE and the truncation trick. In particular, the truncation trick often led to omissions or mispronunciations. Figure~\ref{fig:prosody_mel} illustrates how the spectrogram changes when pitch or duration is controlled. We used Whisper~\cite{radford2023robust} to measure the Phone-Error-Rate (PER), finding that prototype loss plays a crucial role in maintaining speech intelligibility under large prosodic shifts (see Figure~\ref{fig:prosody_control_per}). Additionally, the edits can be localized to specific phonemes or words by selecting only the corresponding encoder activations for adjustment, as shown in the audio examples on our demo webpage.

\subsection{Correction of Mispronunciations}

To correct mispronunciations within synthesized speech, we assume supervision in the form of a speech-only correction query. The premise is that word-level pronunciation information, more easily obtainable than phonetic details, can serve as a practical foundation for corrective adjustments.

Building on the prosody control method, we target mispronunciation correction by manipulating activations in the final Tacotron 2 encoder layer. Our goal is to adjust the predicted semantic tokens for each grapheme to match those of the correction query, ensuring more accurate pronunciation.

To evaluate pronunciation correction, we synthesized 78 speech samples, each with a target word in the carrier sentence ``How is ... pronounced?''. These target words were selected from prior work \cite{fong2022speech} as being outside the LJSpeech training set or challenging for grapheme-input TTS models.

Mispronunciation correction, using Manifold Preserving CAE, was evaluated using four metrics: Word-Error-Rate (WER), Phone-Error-Rate (PER), sentence embedding cosine similarity, and token-based similarity. For cosine similarity, we converted synthesized speech to text using Whisper ASR, then compared sentence embeddings from the text-ada-002 model to ground truth. Token-based similarity was measured using the GPT-2 \cite{radford2019language} tokenizer on Whisper transcripts, assessing the ratio of correctly identified tokens to the ground truth.

\begin{table}[th]
  \centering
  \vspace{-0.1cm}
  \caption{The objective correction capability comparisons.}
  \vspace{-0.2cm}
  \begin{tabular}{@{\hskip 0.03in}c@{\hskip 0.01in}cc@{\hskip 0.03in}}
    \hline
    Metric & Before Correction & After Correction  \\
    \hline
    WER ($\downarrow$) & 0.151 & \textbf{0.056}    \\
    PER ($\downarrow$) & 0.069 & \textbf{0.017}    \\
    Cosine Similarity ($\uparrow$) & 0.954 & \textbf{0.969}    \\
    Token Similarity ($\uparrow$) & 0.892 & \textbf{0.929}    \\
    \hline
    \end{tabular}
    \vspace{-0.3cm}
    \label{tab:sentence_similarity}
\end{table}

As indicated in Table \ref{tab:sentence_similarity}, our CAE approach is effective across all pronunciation correction metrics, substantially reducing both WER and PER, while increasing the semantic similarity of the corrected outputs.

For evaluating the perceptual quality, we conducted a Comparison Mean Opinion Score (CMOS) \cite{loizou2011speech} test comparing initial synthesized speech with corrected speech. Twenty listeners rated the quality, with results presented in Table \ref{tab:cmos}.

\begin{table}[th]
  \centering
  \vspace{-0.1cm}
  \caption{CMOS comparison.}
  \vspace{-0.2cm}
  \begin{tabular}{cc}
    \hline
    Setting             & CMOS      \\
    \hline
    After Correction   & 0.000     \\
    Before Correction    & -0.764    \\
    \hline
    \end{tabular}
    \vspace{-0.3cm}
    \label{tab:cmos}
\end{table}

The results show that correcting prosody and mispronunciations with our CAE approach led to a 0.764-point improvement in CMOS, demonstrating its effectiveness.

\section{Conclusions and Discussion}

\begin{figure}
    \centering
    \includegraphics[width=0.85\linewidth]{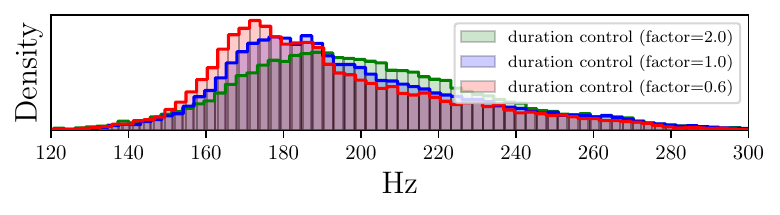}
    \vspace{-0.4cm}
    \caption{Phoneme pitch density over the LJSpeech dataset after Manifold Preserving CAE duration control.}
    \vspace{-0.5cm}
    \label{fig:limitation}
\end{figure}

In this work, we introduced a novel method for prosody control and pronunciation correction in TTS models using Counterfactual Activation Editing, enabling post-hoc adjustments without retraining. Our experiments demonstrated that intermediate representations within TTS models, specifically Tacotron 2, contain rich information that can be effectively manipulated to modify prosody and correct mispronunciations. 

Despite its effectiveness, our approach manipulates entire vectors of intermediate representations, which can overlook individual dimension contributions. For instance, as shown in Figure \ref{fig:limitation}, controlling duration slightly affects pitch, suggesting feature entanglement. Future investigations will aim to refine our approach by targeting specific dimensions (neurons) within these representations. This direction holds the promise of achieving a more granular, interpretable, and disentangled control over speech characteristics.



\section{Acknowledgements}

This work was supported by Samsung Electronics MX Division and by Institute of Information \& communications Technology Planning \& Evaluation (IITP) grant funded by the Korea government (MSIT) (No.2019-0-00075, Artificial Intelligence Graduate School Program (KAIST); No. 2022-0-00984, Development of Artificial Intelligence Technology for Personalized Plug-and-Play Explanation and Verification of Explanation; No. RS-2024-00457882, AI Research Hub Project).



\bibliographystyle{IEEEtran}
\bibliography{mybib}

\end{document}